\documentclass[11pt, a4paper]{article}
\usepackage[margin=1in]{geometry}
\usepackage{amsmath, amssymb, amsfonts}
\usepackage{graphicx}
\usepackage{hyperref}
\usepackage{physics}
\usepackage{bm}
\usepackage{cite}
\usepackage{authblk}
\usepackage{setspace}
\usepackage{abstract}
\usepackage{orcidlink}
\numberwithin{equation}{section}

\title{\textbf{York’s Cavity Formalism and Quantum-Modified Thermodynamics of (2+1)D Black Holes} \\
\vspace{0.4cm}
}

\author[1]{Prince A Ganai, \orcidlink{0000-0002-7896-9183}\thanks{princeganai@nitsri.ac.in}}
\author[1]{Nadeem Ul Islam,\thanks{drnadeemulislam@gmail.com}}

\author[1]{Nazir A Ganaie \thanks{nazirahmadgan.82225@jk.gov.in}}
\affil[1]{\small Department of Physics, National Institute of Technology , Srinagar, India-190006}

\date{\today}

\begin{document}
\maketitle
\begin{abstract}
We explore the canonical thermodynamics of the non-rotating BTZ black hole within a finite cavity by  incorporating quantum corrections using Barrow entropy.  We derive analytic expressions for temperature, quasilocal energy, free energy, and specific heat, all evaluated at the cavity boundary by using York's formalism.  While the redshifted temperature and energy stay the same despite the entropy changes, the altered entropy affects the thermodynamic landscape. Specifically, we see that the Helmholtz free energy drops faster as the horizon size increases. The specific heat shows clear peaks, and their position and height depend on the Barrow parameter. These features signal enhanced thermal responsiveness and a shift in the onset of black hole dominance. Our results demonstrate that quantum entropy corrections alone, without modifying the geometry, can yield rich thermodynamic behaviour in lower-dimensional gravity and highlight the effectiveness of York’s framework in capturing such effects.
\end{abstract}

\section{Introduction}
The thermodynamics of black holes is one of the most important areas for exploring the connection between gravity, quantum theory, and statistical mechanics. Identifying surface gravity with temperature and horizon area with entropy has turned black holes into real thermodynamic objects. They follow laws similar to those of classical thermodynamics\cite{Bekenstein1973,Hawking1976,Gibbons1977,Braden1990}. This deep correspondence, however, prompts further inquiry: What happens to these laws in modified theories of gravity or under quantum corrections to the entropy-area relation?

In lower-dimensional settings such as the \( (2+1) \)-dimensional Banados–Teitelboim–Zanelli (BTZ) black hole \cite{Banados1992}, many features of black hole thermodynamics persist, albeit in a simplified geometric context. 
The BTZ solution has a negative curvature and no curvature singularities. It has been significant in studying boundary conditions, holography, and thermal ensembles in AdS gravity.
Despite the lower dimensionality, the BTZ black hole admits a well defined temperature, entropy, and thermodynamic volume, allowing for meaningful studies of equilibrium and stability\cite{Carlip1995,Huang2022}.

A particularly powerful framework for defining black hole thermodynamics at finite radius is York’s cavity formalism \cite{York1986}. In this approach, the black hole is enclosed within a fixed boundary at finite coordinate radius \( R \), where physical observers reside and all thermodynamic quantities are defined. This strategy avoids divergences at spatial infinity, ensures a well-posed canonical ensemble, and allows quasilocal quantities such as energy and free energy to be evaluated directly from the geometry and boundary data\cite{Brown1993}. While widely applied to four-dimensional spacetimes, relatively fewer studies have employed York's method in three dimensions, especially in the presence of quantum corrections or modified entropy formulations.

One particularly interesting quantum correction comes from the idea of fractal horizons, as proposed by Barrow \cite{Barrow2020,Capozziello2025}.  This idea is driven by quantum gravitational arguments that indicate spacetime near the Planck scale might differ from smooth manifolds. Barrow introduced a phenomenological change to the Bekenstein-Hawking entropy in the following form

\[
S_B = \left( \frac{A}{A_0} \right)^{1 + \Delta/2},
\]
where \( A \) is the horizon area, \( A_0 \) is a constant with units of area, and \( \Delta \) is a dimensionless deformation parameter encoding the degree of quantum-induced roughness. This correction modifies the entropy–area relation without altering the underlying spacetime geometry, thus providing an ideal setting for isolating the effects of entropy deformations on black hole thermodynamics.

The interplay between Barrow entropy and lower dimensional gravity is largely unexplored. While most existing work focuses on asymptotically flat or AdS spacetimes in four or higher dimensions, few analyses have considered how such entropy corrections manifest in a cavity confined \( (2+1) \)D black hole setup. Furthermore, the thermodynamic implications of fractal horizon geometry particularly for stability, phase structure, and heat capacity have not been systematically studied within a quasilocal framework.

In this work, we investigate the canonical thermodynamics of the non-rotating and uncharged BTZ black hole in the presence of Barrow entropy, employing York’s formalism to regulate and define all quantities at a finite radius. Our goal is to quantify how entropy corrections alone, absent any geometric deformation, impact key thermodynamic variables such as free energy and specific heat. By deriving fully analytic expressions for these quantities, we are able to track the influence of the Barrow parameter \( \Delta \) across the full range of physically admissible horizon radii. Special attention is given to the emergence of peaks in specific heat, the shift in free energy zero crossings, and the conditions under which quantum corrected black holes become thermodynamically favored.

The structure of the paper is as follows: In Section II, we construct the thermodynamic ensemble using York’s formalism and derive the redshifted temperature, quasilocal energy, and free energy expressions. In Section III, we introduce Barrow entropy and rederive the modified thermodynamic relations under this correction. Section IV presents a detailed analysis of the resulting thermodynamic behavior, supported by analytic expressions and graphical plots, and concludes with physical interpretations and implications. 

\section{Canonical Thermodynamics via York’s Formalism}
To investigate the thermodynamic properties of the BTZ black hole in a physically meaningful setting, we adopt the quasilocal formalism developed by York \cite{York1986} and extended by Brown and York \cite{Brown1993}. In this framework a canonical ensemble is constructed by enclosing the black hole within a finite, timelike boundary, often referred to as a cavity located at a fixed radial coordinate \( R \), with \( R > r_+ \). All thermodynamic quantities are then defined relative to this boundary, enabling the formulation of black hole thermodynamics at finite radius without requiring access to asymptotic infinity.

\subsection *{BTZ Geometry and Cavity Setup}

The non-rotating and uncharged BTZ black hole metric given by 
\begin{equation}
ds^2 = -f(r)\,dt^2 + \frac{dr^2}{f(r)} + r^2\,d\phi^2, \qquad f(r) = -M + \frac{r^2}{l^2},
\label{eq:BTZmetric}
\end{equation}
where \( M \) is the ADM mass and \( l \) is the AdS curvature radius, which is related to the cosmological constant by \( \Lambda = -1/l^2 \). Event horizon lies at \( r = r_+ \), satisfying \( f(r_+) = 0 \), which gives
\begin{equation}
r_+ = l\sqrt{M}.
\label{eq:horizon}
\end{equation}

We introduce a circular boundary at fixed radius \( r = R \) in order to formulate thermodynamics in a cavity, with the induced metric
\[
ds^2_{\text{boundary}} = -f(R)\,dt^2 + R^2\,d\phi^2.
\]

For an observer the boundary is in the local frame for evaluating thermodynamic quantities.

\subsection*{Redshifted Temperature}
The temperature measured locally at the cavity wall differs from the Hawking temperature due to gravitational redshift phenomena. By Tolman's law,  the redshifted  temperature is defined as follows
\begin{align}
    T(R) = \frac{T_H}{\sqrt{f(R)}},\\
\text{where} \quad T_H = \frac{1}{4\pi} \left. \frac{df}{dr} \right|_{r = r_+} 
\nonumber
\end{align}
Differentiating the metric function  \( f'(r) = \frac{2r}{l^2} \), will give us  Hawking temperature 
\begin{equation}
T_H = \frac{r_+}{2\pi l^2}
\end{equation}

Substituting into Tolman’s formula, and noting that \( f(R) = \frac{R^2 - r_+^2}{l^2} \), we obtain 
\begin{equation}
T(R) = \frac{r_+}{2\pi l \sqrt{R^2 - r_+^2}}.
\label{eq:T_R}
\end{equation}
\noindent
To ensure that the redshifted temperature remains finite and the observer lies outside the black hole horizon, we assume \( R > r_+ \) throughout the analysis.
\subsection*{Brown–York Quasilocal Energy and Background Subtraction}

The energy of the system, as seen  by an observer at the cavity boundary, is defined using the Brown-York quasilocal prescription given as 
\begin{equation}
E(R) = \frac{1}{8\pi} \left[ k(R) - k_0(R) \right] \cdot \mathcal{A}(R),
\label{eq:BY_E}
\end{equation}
where \( \mathcal{A}(R) = 2\pi R \) is the circumference of the boundary, \( k \)  is the trace of the extrinsic curvature of the boundary embedded in the black hole spacetime, and \( k_0 \) is the corresponding quantity computed in a reference background. The subtraction ensures that the energy is finite and normalised to vanish without a black hole.

We consider following the spacelike unit normal to the boundary to compute  \( k(R) \)
\[
n^\mu  = \left(0,  \sqrt{f(R)}, 0\right),
\]
And apply:
\begin{equation}
k(R) = \nabla_\mu n^\mu = \frac{f'(R)}{2\sqrt{f(R)}} = \frac{R}{l^2 \sqrt{f(R)}} = \frac{R}{l \sqrt{R^2 - r_+^2}}.
\end{equation}

For the reference geometry, we choose pure AdS space (\( M = 0 \)), with metric function \( f_{\text{AdS}}(r) = \frac{r^2}{l^2} \), resulting
\begin{equation}
k_0(R) = \frac{f_{\text{AdS}}'(R)}{2\sqrt{f_{\text{AdS}}(R)}} = \frac{1}{l}.
\label{eq:k0}
\end{equation}

Substituting \( k(R) \) and \( k_0(R) \) into Eq.~\eqref{eq:BY_E}, we find:
\begin{align}
E(R) &= \frac{1}{8\pi} \left( \frac{R}{l \sqrt{R^2 - r_+^2}} - \frac{1}{l} \right) \cdot 2\pi R \nonumber \\
     &= \frac{R}{4l} \left( \frac{R}{\sqrt{R^2 - r_+^2}} - 1 \right).
\label{eq:E_R}
\end{align}

Background subtraction is necessary. Without the reference term \( k_0 \), the energy would diverge as \( R \to \infty \),  since the AdS background has infinite total energy. Computing energy relative to AdS, we ensure that \( E(R) \to 0 \) as \( r_+ \to 0 \), is consistent with the interpretation that no black hole corresponds to the vacuum. This reflects the general principle in general relativity that energy is only meaningful as a relative quantity, evaluated between spacetimes with the same boundary data.

\subsection*{Free Energy in the Canonical Ensemble}

The Helmholtz free energy is defined by:
\begin{equation}
F(R) = E(R) - T(R)\,S,
\end{equation}
where \( S \) is the black hole entropy. For the BTZ black hole, the entropy follows from the Bekenstein–Hawking formula:
\begin{equation}
S = \frac{A}{4} = \frac{\pi r_+}{2}.
\end{equation}
Substituting \( T(R) \), \( E(R) \), and \( S \) into the free energy expression, we obtain:
\begin{equation}
F(R) = \frac{R}{4l} \left( \frac{R}{\sqrt{R^2 - r_+^2}} - 1 \right) - \left( \frac{r_+}{2\pi l \sqrt{R^2 - r_+^2}} \right) \cdot \left( \frac{\pi r_+}{2} \right).
\label{eq:F_R}
\end{equation}

This expression, which depends explicitly on the horizon radius and the cavity size, serves as the central thermodynamic potential for the canonical ensemble. The next section will analyse how quantum corrections, introduced through generalised entropy models, modify this structure and influence thermodynamic behaviour.

\section{Quantum Corrections via Barrow Entropy}

Quantum gravitational effects are expected to modify the standard entropy area relation of black holes. One intriguing model of such corrections is Barrow entropy\footnote {Barrow proposed that quantum gravitational effects may induce a fractal structure on the black hole horizon, modifying the standard area law. In this framework, the entropy scales non-linearly with horizon length (or area in higher dimensions), leading to the form \( S_B \propto A^{1 + \Delta/2} \) where \( \Delta \in [0,1] \) measures the deviation from smooth geometry \cite{Barrow2020}.}, which arises from the hypothesis that spacetime near a black hole horizon is not smooth but instead exhibits quantum fractal features. This leads to a deformation of the Bekenstein–Hawking entropy into a non-extensive form:

\begin{equation}
S(r_+) = \left( \frac{\pi r_+}{2} \right)^{1 + \Delta/2},
\label{eq:barrow_entropy_def}
\end{equation}
\noindent
While the semiclassical approximation remains valid for small deformation parameters \( \Delta \), it is worth noting that for \( \Delta \sim 1 \), backreaction effects may become significant. In such regimes, the black hole geometry itself could receive quantum corrections, which lie beyond the scope of this study. Our analysis is thus restricted to cases where \( \Delta \) remains within the semiclassical regime.

where \( \Delta \in [0, 1] \) is the Barrow deformation parameter, and \( r_+ \) is the horizon radius. The classical result is recovered in the limit \( \Delta \to 0 \). This modification does not affect the spacetime geometry, allowing us to work within the standard BTZ metric and York’s cavity formalism while adjusting the thermodynamic framework solely through the entropy.

In Barrow’s original formulation, the parameter \( \Delta \) lies in the interval \( 0 \leq \Delta \leq 1 \), ensuring that the entropy remains finite and sub-extensive. The limit \( \Delta = 0 \) recovers the classical Bekenstein–Hawking result, while \( \Delta = 1 \) corresponds to a maximally fractal horizon surface. Observational constraints on \( \Delta \) are not yet well established, but potential bounds could emerge from holographic considerations or  gravitational wave tests of horizon thermodynamics. In our analysis, we work within the theoretically motivated range \( \Delta \in [0, 1] \), and examine how increasing \( \Delta \) affects thermodynamic observables. A more refined constraint could arise from entropy bounds, holographic entanglement entropy, or corrections to black hole mergers, which are intriguing  directions for future exploration.

We adopt the  standard semiclassical assumption that quantum gravitational effects manifest dominantly through entropy deformation while the spacetime geometry remains classical. This is supported by the fact that  Barrow entropy originates from horizon microstructure, not from modifications to the field equations themselves \cite{Barrow2020}. Moreover, in (2+1) dimensions, the BTZ metric solves the  Einstein equations with a negative cosmological constant and no local degrees of freedom, making the assumption of a fixed metric particularly reasonable. Nevertheless, in a full quantum  gravity framework, backreaction effects could potentially modify the geometry. In such a case, the metric function \( f(r) \) and consequently the  redshifted temperature \( T(R) \) may receive higher order corrections. Investigating these effects would require solving the modified Einstein equations with an effective stress energy tensor sourced by the fractal deformation, which lies beyond the scope of this work.

\subsection*{Modified Helmholtz Free Energy}

In the canonical ensemble at fixed boundary radius \( R \), the Helmholtz free energy is defined by

\begin{equation}
F(R) = E(R) - T(R)\,S(r_+),
\end{equation}

where \( E(R) \) is the Brown–York quasilocal energy and \( T(R) \) is the redshifted temperature, both of which remain unchanged from the classical setup:

\begin{align}
E(R) &= \frac{R}{4l} \left( \frac{R}{\sqrt{R^2 - r_+^2}} - 1 \right), \\
T(R) &= \frac{r_+}{2\pi l \sqrt{R^2 - r_+^2}}.
\end{align}

Substituting Eq.~\eqref{eq:barrow_entropy_def} into the definition of free energy yields:

\begin{align}
F(R) = \frac{R}{4l} \left( \frac{R}{\sqrt{R^2 - r_+^2}} - 1 \right)
- \frac{r_+}{2\pi l \sqrt{R^2 - r_+^2}} \cdot \left( \frac{\pi r_+}{2} \right)^{1 + \Delta/2}.
\label{eq:free_energy_barrow}
\end{align}
In the limit \( \Delta \rightarrow 0 \), Eq.~\eqref{eq:free_energy_barrow} smoothly reduces to the classical Helmholtz free energy of the BTZ black hole in a cavity, confirming that the deformation respects the expected semiclassical limit.
This expression demonstrates how the entropy deformation introduces a nontrivial \( r_+ \)-dependence in the free energy, even though the geometry remains unchanged.

\subsection*{Specific Heat at Fixed Radius}

To determine the thermodynamic stability of the system, we evaluate the specific heat at fixed cavity radius \( R \), defined by

\begin{equation}
C_R = T(R)\left( \frac{\partial S}{\partial T(R)} \right)_R = T(R)\frac{dS/dr_+}{dT(R)/dr_+}.
\end{equation}

Differentiating Eq.~\eqref{eq:barrow_entropy_def} with respect to \( r_+ \) gives:

\begin{equation}
\frac{dS}{dr_+} = \left( 1 + \frac{\Delta}{2} \right) \left( \frac{\pi}{2} \right)^{1 + \Delta/2} r_+^{\Delta/2}.
\end{equation}

The redshifted temperature is:

\[
T(R) = \frac{r_+}{2\pi l \sqrt{R^2 - r_+^2}},
\]

and differentiating with respect to \( r_+ \) gives:

\begin{equation}
\frac{dT(R)}{dr_+} = \frac{1}{2\pi l \sqrt{R^2 - r_+^2}} + \frac{r_+^2}{2\pi l (R^2 - r_+^2)^{3/2}}
\end{equation}

Combining, the specific heat becomes:

\begin{equation}
C_R = \left( 1 + \frac{\Delta}{2} \right)
\left( \frac{\pi}{2} \right)^{1 + \Delta/2}
\cdot \frac{r_+^{1 + \Delta/2}(R^2 - r_+^2)}{R^2}.
\label{eq:final_specific_heat}
\end{equation}

It is worth emphasizing that \( C_R \rightarrow 0 \) as \( r_+ \rightarrow R \), which is expected since the redshifted temperature diverges and the black hole horizon approaches the cavity wall. This ensures thermodynamic regularity and consistency within York’s framework.

The result is manifestly positive and regular for all \( 0 < r_+ < R \), ensuring local thermodynamic stability of the black hole inside the cavity. As \( \Delta \to 0 \), one recovers the classical BTZ specific heat.

Notably, the form of Eq.~\eqref{eq:final_specific_heat} shows that quantum corrections introduced solely through the entropy can significantly alter the heat capacity profile. This will be further discussed in Section IV through both analytical trends and graphical results.

\begin{figure}[htbp]
\centering
\includegraphics[width=\linewidth]{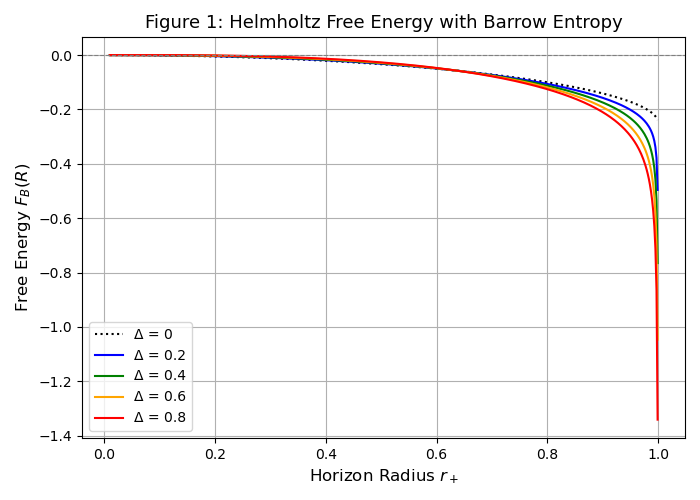} 
\caption{Helmholtz free energy $F_B(R)$ versus horizon radius $r_+$ for various Barrow deformation parameters $\Delta$. The classical case $\Delta = 0$ is shown as a dotted black curve.}
\label{fig:free_energy}
\end{figure}

\begin{figure}[htbp]
\centering
\includegraphics[width=\linewidth]{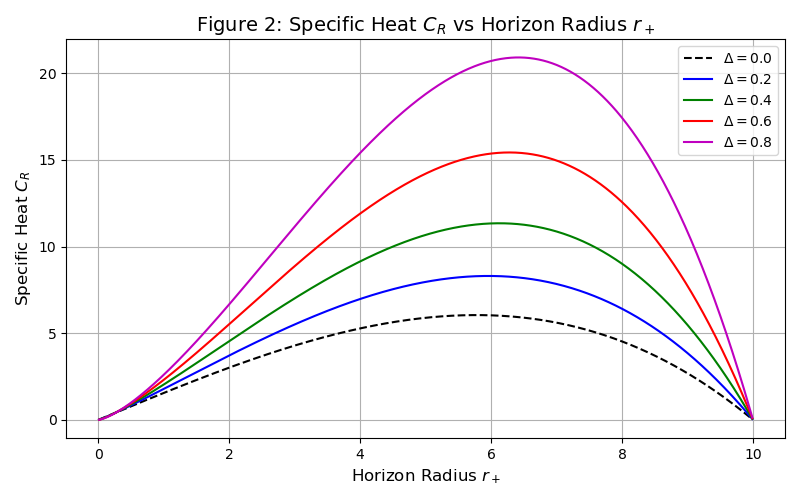}
\caption{Specific heat $C_R$ as a function of horizon radius $r_+$ for several Barrow deformation parameters $\Delta$. The classical BTZ case ($\Delta = 0$) is shown as a dotted black line. The increase in $C_R$ with $\Delta$ indicates enhanced thermodynamic stability due to quantum corrections.}
\label{fig:specific_heat}
\end{figure}

\subsection*{Comment on Equation of State}

Although the extended phase-space formalism, in which the cosmological constant is treated as thermodynamic pressure, is not invoked because of the fixed cavity setup, the Barrow deformation introduces effects analogous to a modified equation of state. In particular, the altered scaling of the entropy effectively changes the system’s heat capacity and the free energy landscape, much like how pressure-volume dynamics reshape the phase behavior in black hole chemistry. In this sense, Barrow entropy plays a role similar to a microstructural deformation of the thermodynamic state space, affecting stability and equilibrium transitions without the need to vary the bulk spacetime parameters.

\section{Results, Discussion and Conclusion}

Figures~1 and 2 encapsulate the thermodynamic response of the BTZ black hole when confined to a finite cavity and corrected via Barrow’s entropy model. Since the spacetime geometry remains unmodified, the influence of quantum deformation enters solely through the entropy function, providing a clean framework for isolating quantum microstructural effects.

The Helmholtz free energy \( F(R) \), shown in Figure~1, reflects the thermodynamic cost of sustaining a black hole of radius \( r_+ \) within a cavity held at fixed boundary temperature. As the Barrow deformation parameter \( \Delta \) increases, the entropy contribution to \( F = E - T S \) grows more rapidly, leading to a sharper decline in free energy with horizon size. The point where \( F(R) \) crosses zero shifts toward smaller \( r_+ \), indicating that quantum-deformed black holes become thermodynamically favorable at smaller sizes. This shift marks a transition from a thermal AdS-like phase to a black hole-dominated phase and emerges entirely from entropy deformation, not geometric change.
\noindent
Even for \( \Delta = 1 \), the free energy remains well-behaved, though it decays more rapidly due to the enhanced entropy growth. No divergences or unphysical behavior arise within the range \( r_+ < R \).

Figure~2 reveals a more intricate pattern in the specific heat \( C_R \), which measures the black hole’s sensitivity to temperature fluctuations within the cavity. For all values of \( \Delta \), including the classical case of \(\Delta=0\),  \( C_R \) increases from near zero, peaks at an intermediate \( r_+ \), and then decreases as the horizon approaches the cavity wall. These peaks signify regions of maximal thermal responsiveness, where small temperature changes lead to substantial entropy variation. In finite systems, such non-monotonic behavior often signifies a smooth crossover in thermodynamic regimes, analogous to what occurs near critical points in extended systems.
\noindent
The peaks in \( C_R \) suggest a regime where the system’s thermal fluctuations are maximally coupled to the underlying microstructure introduced by Barrow deformation. This behavior resembles the appearance of soft modes in statistical systems near criticality, albeit without an actual phase transition. Such peaks reflect a nontrivial restructuring of the entropy–temperature relation and mark a thermodynamically privileged scale within the cavity.

As \( \Delta \) increases, the peaks in \( C_R \) grow more pronounced and shift towards  larger radii. This behaviour indicates that the Barrow deformation improves the system’s capacity to absorb energy at specific horizon scales, effectively introducing a new thermodynamic structure without modifying the bulk geometry. This effect stems from the non-extensive, fractal character of the modified entropy and highlights how microstructure at the horizon can influence macroscopic observables.

Throughout the physical range of horizon radius \( 0 < r_+ < R \), the specific heat remains positive, indicating  that the black hole configurations are locally thermodynamically stable. The analytic expressions for both \( F(R) \) and \( C_R \) reduce smoothly to their classical counterparts in the limit \( \Delta \to 0 \), confirming the internal consistency of the deformation model. These results affirm that Barrow entropy introduces meaningful corrections even in lower-dimensional gravity, which lacks local dynamical degrees of freedom.

In summary, entropy-based quantum corrections can fundamentally reshape black hole thermodynamics without requiring geometric backreaction. The results presented here, earlier thermodynamic preference, enhanced thermal capacity, and structured heat response, demonstrate that modifications to horizon microstructure alone can leave observable imprints. York’s cavity framework offers a particularly effective setting for examining such corrections, as it allows precise boundary-based definitions of temperature, energy, and stability. We anticipate that these findings may serve as a stepping stone toward incorporating more general entropy models and exploring their implications in a wide range of gravitational settings.




\end{document}